\documentclass[twocolumn]{revtex4}
\usepackage{graphicx}

\begin{document}

\title{Magnetic Properties of J-J-J$^{\prime}$ Quantum Heisenberg Chains
with Spin $S=1/2$, $1$, $3/2$ and $2$ in a Magnetic Field}
\author{ Bo Gu and Gang Su$^{\ast }$ }
\affiliation{College of Physical Sciences, Graduate University of Chinese Academy of
Sciences, P. O. Box 4588, Beijing 100049, China}
\author{Song Gao}
\affiliation{College of Chemistry and Molecular Engineering, State Key Laboratory of Rare
Earth Materials Chemistry and Applications, Peking University, Beijing
100871, China}

\begin{abstract}
By means of the density matrix renormalization group (DMRG) method, the
magnetic properties of the J-J-J$^{\prime }$ quantum Heisenberg chains with
spin $S=1/2$, $1$, $3/2$ and $2$ in the ground states are investigated in
the presence of a magnetic field. Two different cases are considered: (a)
when $J$ is antiferromagnetic and $J^{\prime }$ is ferromagnetic (i.e. the
AF-AF-F chain), the system is a ferrimagnet. The plateaus of the
magnetization are observed. It is found that the width of the plateaus
decreases with increasing the ferromagnetic coupling, and disappears when $%
J^{\prime }/J$ passes over a critical value. The saturated field is observed
to be independent of the ferromagnetic coupling; (b) when $J$ is
ferromagnetic and $J^{\prime }$ is antiferromagnetic (i.e. the F-F-AF
chain), the system becomes an antiferromagnet. The plateaus of the
magnetization are also seen. The width of the plateaus decreases with
decreasing the antiferromagnetic coupling, and disappears when $J^{\prime
}/J $ passes over a critical value. Though the ground state properties are
quite different, the magnetization plateaus in both cases tend to disappear
when the ferromagnetic coupling becomes more dominant. Besides, no
fundamental difference between the systems with spin half-integer and
integer has been found.

\noindent PACS number(s): 75.10.Jm, 75.30.Kz
\end{abstract}

\maketitle

\section{INTRODUCTION}

Low-dimensional quantum spin systems have been attracting both experimental
and theoretical interest in the last decades due to an interplay of strong
quantum fluctuations and topology. Several theoretical predictions for the
low-dimensional quantum spin chains have been verified by experimental
studies. For the Heisenberg antiferromagnetic (HAF) chains with spin $%
S=half-integer$, the celebrated Lieb, Schultz and Mattis theorem showed that
the excitation of the system is gapless \cite{LSM}. For the HAF chains with
spin $S=integer$, the excitation from the singlet ground state to the
triplet excited state was conjectured to be gapful, now known as Haldane
conjecture\cite{Haldane}. Another interesting phenomenon in HAF spin chains
is the occurrence of the magnetization plateaus, that can be viewed as an
essentially macroscopic quantum phenomenon, and has gained much attention
recently. A decade ago, Hida has considered an $S=1/2$ HAF chain with
exchange coupling of $3$-site translational invariance in the presence of an
applied magnetic field, and uncovered a plateau in the magnetization curve
at $1/3$ of the saturation magnetization\cite{Hida}. Slightly after Hida's
numerical calculation on the plateau of the $S=1/2$ F-F-AF chain, an
analytical approach was done by Okamoto\cite{Okamoto}. These results lead to
a more general necessary condition for the appearance of the magnetization
plateaus which was proved by Oshikawa, Yamanaka and Affleck\cite{OYA}. It
tells us that for the HAF spin chains with $S=integer$ or $half-integer$,
the magnetization curve can have plateaus at which the magnetization per
site $m$ is topologically quantized by 
\begin{equation}
n(S-m)=integer,  \label{plateau}
\end{equation}%
where $S$ is the magnitude of the spin, and $n$ is the period of the ground
state determined by the explicit spatial structure of Hamiltonian. At the
plateaus, the spin gaps open, that can be in some sense regarded as a kind
of generalization of the Haldane conjecture. Similar to the quantum Hall
effect, the magnetization plateau is another striking example of the
macroscopic quantum phenomenon, in which magnetization is quantized to
fractional values of the saturated magnetization value and is a function of
the magnetic field.

The magnetization plateaus are predicted and observed in many
low-dimensional spin systems. Among others, a magnetization plateau at half
the saturation magnetization was observed in $S=1$ HAF bond-alternating
chain compounds $\mathrm{[Ni_{2}(dpt)_{2}(\mu -ox)(\mu -N_{3})](PF_{6})} $
(abbreviated as NDOAP) and $\mathrm{[Ni(333-tet)(\mu -NO_{2})]ClO_{4} }$
(abbreviated as NTENP), where the experimental result is in agreement with
the numerical calculations\cite{HABA1,HABA2,HABA3}; the ferrimagnetic mixed
spin chains such as the bimetallic chain $\mathrm{MM^{\prime
}(pbaOH)(H_{2}O)nH_{2}O}$ and the organic compound $\mathrm{{Mn(hfas)_{2}}%
_{3}(3R)_{2}}$ show quantum magnetization plateaus\cite%
{FERRI1,FERRI2,FERRI3, FERRI4,FERRI5}; the magnetization plateaus are also
found in p-merized chains and ladders\cite%
{p-chain1,p-chain2,p-chain3,p-chain4}. The effect of randomness on
magnetization process also attracts much theoretical interest \cite%
{Random1,Random2}. One-dimensional (1D) helical spin system $\mathrm{%
Co(hfac)_{2}NITPhOMe}$ shows some interesting magnetic behaviors, where one
of the unusual properties is that the magnetization shows plateaus at zero
and $\frac{1}{3}$ of the saturation if a magnetic field is applied along the
helical axis, but no plateaus if the field is applied in the plane
perpendicular to that axis\cite{Helical}. Another intriguing topic in 1D
spin systems is concerned with the spin-orbital model, such as $\mathrm{%
Na_{2}Ti_{2}Sb_{2}O }$ and $\mathrm{NaV_{2}O_{5}}$, where the orbital degree
of freedom plays an important role in the appearance of plateaus \cite%
{spin-orbit1,spin-orbit2,spin-orbit3,spin-orbit4}.

Despite the existence of considerable theoretical insight, a satisfactory
understanding of the experimental situation is still sparse. Experimentally,
polynuclear azido-bridged derivatives are a rich source of new magnetic
systems. Because of the extreme versatility of the coordination modes of the
azido ligand, these magnetic systems may coordinate as $\mathrm{\mu
_{1,3}-N_{3}}$(end-to-end, EE), $\mathrm{\mu _{1,1}-N_{3}}$(end-on, EO), or
even in more exotic modes as $\mathrm{\mu _{1,1,1}-N_{3}}$ or $\mathrm{\mu
_{1,1,3}-N_{3}}$ \cite{AFAFF1,AFAFF2,AFAFF3}. The similar degree of
stability of the EE or EO coordination modes often favors a variety of
topologies or dimensionalities. The antiferromagnetic (AF) interaction and
the ferromagnetic (F) coupling are generally found in the EE and EO mode,
respectively. The exotic topologies with alternating patterns EE/EE/EO have
been reported \cite{AFAFF11}, which offers exciting research prospects.

In this paper, motivated by the exotic magnetic properties of the period $%
n=3 $ quantum spin system such as EE/EE/EO, we shall investigate the ground
state of a 1D J-J-J$^{\prime }$ quantum spin chain with spin $S=1/2$, $1$, $%
3/2$, and $2$, respectively. It is found that in the magnetic process,
magnetization plateaus are observed, in agreement with Eq.(\ref{plateau}).
The width of the magnetization plateaus is found to depend on the ratio of $%
J^{\prime }/J$. The fundamental difference of the properties between the
systems with spin half-integer and integer is not observed.

The rest of this paper is outlined as follows. In Sec. II, we introduce the
model Hamiltonian for the 1D J-J-J$^{\prime }$ Heisenberg spin chain. In
Sec. III, we present our numerical results of the ground state of the
system. A brief summary is given in Sec. IV.

\section{MODEL}

\begin{figure}[tbp]
\includegraphics[width = 0.75\linewidth,clip]{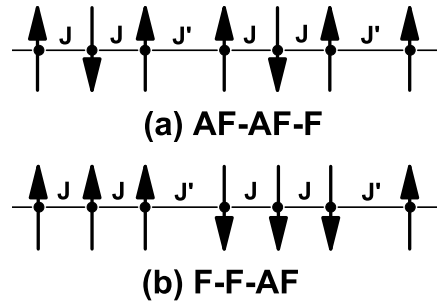}
\caption{ The spin arrangements of the J-J-J$^{\prime }$ chain, where $%
\uparrow $ denotes spin up, $\downarrow $ denotes spin down, $J$ and $%
J^{\prime }$ are exchange couplings. (a) AF-AF-F chain, $J = J_{\rm AF}$, $%
J^{\prime} = J_{\rm F}$; (b) F-F-AF chain, $J = J_{\rm F}$, $J^{\prime } = J_{\rm AF}$. }
\label{chain}
\end{figure}

Motivated by the experimental study of the magnetic coordinated compounds%
\cite{AFAFF1,AFAFF2,AFAFF3}, let us investigate the 1D J-J-J$^{\prime }$
Heisenberg quantum spin system. The Hamiltonian of the system reads 
\begin{equation}
H=\sum\limits_{j}(J\mathbf{S}_{3j-2}\cdot \mathbf{S}_{3j-1}+J\mathbf{S}%
_{3j-1}\cdot \mathbf{S}_{3j}+J^{\prime }\mathbf{S}_{3j}\cdot \mathbf{S}%
_{3j+1})-h\sum\limits_{j}S_{j}^{z},  \label{Hamilton}
\end{equation}%
where $J$ and $J^{\prime }$ are exchange integrals with $J$, $J^{\prime }$ $%
>0$ denoting the AF coupling and $J$, $J^{\prime }<0$ the F coupling, $h$ is
the external magnetic field, and we take $g\mu _{B}=1$. The schematic spin
arrangement of the system is shown in Fig. \ref{chain}.

When $J = J^{\prime } = J_{\rm AF} $, the system becomes a uniform HAF spin
chain. In this case, it is well-known that for $S = half-integer$, there is
no magnetization plateau in the magnetic process before saturation; for $S =
integer$, owing to the existence of the Haldane gap there appears a plateau
at zero magnetization below the lower critical field where the spin gap
closes. When $J = J^{\prime } = J_{\rm F} $, the system becomes a uniform
Heisenberg ferromagnetic chain. It has no magnetization plateaus before
saturation in this case.

When $J \neq J^{\prime }$, the system shows complex behaviors. As indicated
in Fig. \ref{chain}, there are two interesting cases: (a) $J = J_{\rm AF}$, $%
J^{\prime} = J_{\rm F}$. It is a ferrimagnetic spin chain (i.e. the
configuration is AF-AF-F). (b) $J = J_{\rm F} $, $J^{\prime} = J_{\rm AF} $. It is
an antiferromagnetic system (i.e. the configuration is F-F-AF). In order to
probe the fundamental difference of properties of the systems with spin
half-odd integer and integer during the magnetizing process, we shall
consider $S=1/2$, $1$, $3/2$ and $2$ for each case below.

\section{DMRG RESULTS}

The density matrix renormalization group (DMRG)\cite{White} was proposed
more than ten years ago, which nowadays becomes a powerful numerical method
invoked to study the ground state and low-lying states of low-dimensional
lattice systems. In the following, the magnetic properties of the ground
state of the spin chain with configurations (a) and (b) in Fig. \ref{chain}
with open boundary conditions are investigated by the DMRG method. In our
calculations, we took the chain length $L=60$, the number of states kept per
block $N=60$ for spin $S=1/2$ and $1$, and $N=80$ for spin $S=3/2$ and $2$.
The truncation error is less than $10^{-3}$ in all cases.

\subsection{AF-AF-F CHAIN}

Let us first consider the AF-AF-F chain, as shown in Fig. \ref{chain}(a). In
this case, $J = J_{\rm AF}$ and $J^{\prime } = J_{\rm F}$. Obviously, it is a
ferrimagnetic chain. When $J_{\rm F}/J_{\rm AF}$ is much less than $1$, the
antiferromagnetic coupling is dominant; while $J_{\rm F}/J_{\rm AF}$ is much greater
than $1$, the ferromagnetic coupling becomes dominant. Without loss of
generality, we shall take the ratio $J_{\rm F}/J_{\rm AF}$ in the range of $[0.1,
10] $ in this subsection.

\begin{figure}[tbp]
\includegraphics[width = 0.75\linewidth,clip]{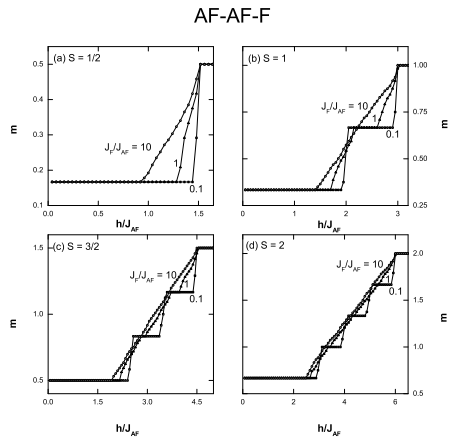}
\caption{Magnetic curves of the AF-AF-F chain of $S=1/2$, $1$, $3/2$ and $2$%
, respectively, where $h$ is the applied magnetic field, $m$ is the
magnetization per site. $J = J_{\rm AF} = 1$, $J^{\prime } = J_{\rm F}$. }
\label{aaf-mh}
\end{figure}

Fig. \ref{aaf-mh} show the magnetization process of the AF-AF-F chain with
spin $S=1/2$, $1$, $3/2$ and $2$, respectively. For $S=1/2$, the two
plateaus at $m=1/6$ and $m=1/2$ (saturation plateau) are obtained,
consistent with the necessary condition given by Eq.(\ref{plateau}) with $%
n=3 $. For $S=1$, the three plateaus at $m=1/3$, $2/3$ and $1$ (saturation
plateau) are seen. For $S=3/2$, the four plateaus at $m=1/2$, $5/6$, $7/6$
and $3/2$ (saturation plateau) are observed, and for $S=2$, the five
plateaus at $m=2/3$, $1$, $4/3$, $5/3$ and $2$ are obtained. The number of
the plateaus is $(2S+1)$. It is clear that the smaller the ratio $%
J_{\rm F}/J_{\rm AF}$ is, the more obvious the plateaus are, that appears to be
independent of the magnitude of spin. In other words, when the AF
interaction in the system plays a dominant role, the magnetization plateau
becomes wider; when the ferromagnetic interaction is dominant, the width of
the plateau gets narrower, as indicated in Fig. \ref{aaf-mh}. In between the
plateaus, the magnetization per site increases with increasing the field for
all cases with different spins. In this system, the saturated field does not
change with the ratio $J_{\rm F}/J_{\rm AF}$. A further discussion on this point can
be found in Sec. IV. For the present system with different spins $S=1/2$, $1$%
, $3/2$ and $2$, the qualitatively similar behaviors of the magnetization
process are observed.

\begin{figure}[tbp]
\includegraphics[width = 0.75\linewidth,clip]{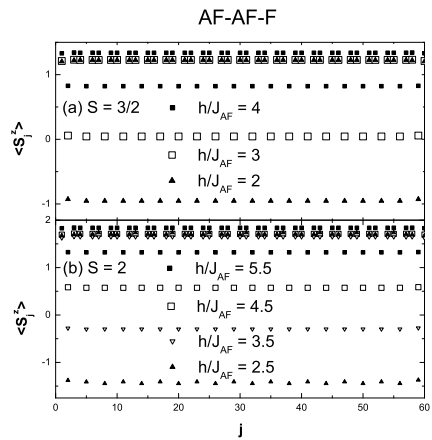}
\caption{ The spatial variation of $\langle S_{j}^{z}\rangle$ in the ground state for the
finite AF-AF-F chain with length$L=60$ at $J_{\rm F}/J_{\rm AF}=0.1$. (a) For $S=3/2$%
, the external field is taken as $h/J_{\rm AF}=2$, $3$, $4$, respectively. (b)
For $S=2$, the external field is taken as $h/J_{\rm AF}=2.5$, $3.5$, $4.5$ and $%
5.5$, respectively. }
\label{aaf-szj}
\end{figure}

The appearance of the magnetization plateaus can be understood from the
spatial dependence of the averaged local magnetic moment, as discussed in
Ref.\cite{hida-affleck}. Fig. \ref{aaf-szj} presents the spatial variation
of $\langle S_{j}^{z}\rangle$ in the ground state for the AF-AF-F chain with spin $S=3/2$
and $2$, respectively. For both cases, the ratio $J_{\rm F}/J_{\rm AF}=0.1$ is
taken. Fig. \ref{aaf-szj}(a) shows the case with $S=3/2$. At the external
field $h/J_{\rm AF}=2$, the expectation values $\langle S_{j}^{z}\rangle$ versus site $j$
follow the sequence such that $\{...,(1.2268,1.2268,-0.9535),...\}$,
resulting in the magnetization per site $m=%
\sum_{j=1}^{N}(\langle S_{3j-2}^{z} \rangle+\langle S_{3j-1}^{z} \rangle
+\langle S_{3j}^{z}\rangle)/3N=1/2$,
corresponding to the plateau $m=1/2$. When the field increases to $h/J_{\rm AF}=3
$, the behavior of $\langle S_{j}^{z}\rangle$ becomes $\{...,(1.2280,1.2280,0.0440),...\}$%
, giving rise to $m=5/6$; at $h/J_{\rm AF}=4$, $\langle S_{j}^{z}\rangle$ versus $j$
behaviors as $\{...,(1.3390,1.3390,0.8219),...\}$, leading to $m=7/6$. Fig. %
\ref{aaf-szj}(b) shows the case with $S=2$. Similarly, at $h/J_{\rm AF}=2.5$,
the expectation value $\langle S_{j}^{z}\rangle$ versus site $j$ behaviors as  $%
\{...,(1.7178,1.7241,-1.4448),...\}$, resulting in $m=2/3$; at $h/J_{\rm AF}=3.5$%
, $\langle S_{j}^{z}\rangle$ varies according to $\{...,(1.6478,1.6478,-0.2956),...\}$,
giving the $m=1$; at $h/J_{\rm AF}=4.5$, $\langle S_{j}^{z}\rangle$ behaviors as $%
\{...,(1.7137,1.7127,0.5746),...\}$, leading to $m=4/3$; and at $h/J_{\rm AF}=5.5
$, it becomes $\{...,(1.8390,1.8390,1.3220),...\}$, giving rise to $m=5/3$.
These $m$ values just correspond to the magnetization plateaus.
In addition, the spin configuration for the parameters above are stable
against the small local field $h_{\rm loc}/J_{\rm AF}$=$0.1$, which is set at
$j=1$ site and parallel to the external magnetic field $h$.

\begin{figure}[tbp]
\vspace{0.5cm} \includegraphics[width = 0.75\linewidth,clip]{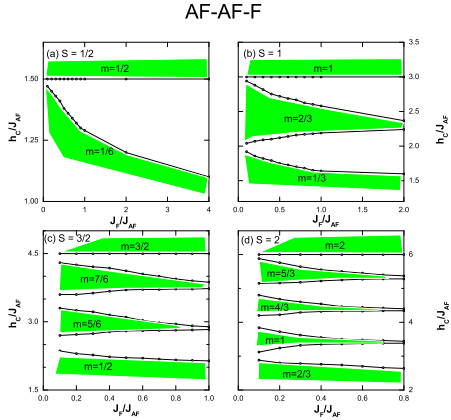}
\caption{ (Color online) Phase diagram in the $h_{\rm c}-J_{\rm F}/J_{\rm AF}$ plane for
the AF-AF-F chain with $S=1/2$, $1$, $3/2$ and $2$, respectively. In the
shaded regions, the magnetization plateaus labeled with the quantized values
of $m$ appear.}
\label{aaf-hc}
\end{figure}

Fig. \ref{aaf-hc} illustrates the phase diagram in the $h_{\rm c}-J_{\rm F}/J_{\rm AF}$
plane for the AF-AF-F chain with spin $S=1/2$, $1$, $3/2$ and $2$,
respectively. In the shaded regions, the magnetization plateaus appear,
implying that the excitations from the ground state are gapful. When $%
J_{\rm F}/J_{\rm AF}$ increases beyond the critical value $(J_{\rm F}/J_{\rm AF})_{\rm c}$ at
which the plateau vanishes, i.e. the ferromagnetic coupling becomes more
dominant, the magnetization plateaus tend to disappear. The
plateau-non-plateau transition could be of the
Berezinskii-Kosterlitz-Thouless(BKT)-type, and the width of the plateau $%
m=\Delta h_{\rm c}$ near the transition point behaves as 
\begin{equation}
\mathrm{\Delta h_{\it c}=Dexp(-C/\sqrt{(J_{\it F}/J_{\it AF})_{\it c}
        -J_{\it F}/J_{\it AF}}) },
\label{aaf-BKT}
\end{equation}%
where $D,C$ are constants\cite{FFACRI}.

\begin{figure}[tbp]
\vspace{0.5cm} \includegraphics[width = 0.75\linewidth,clip]{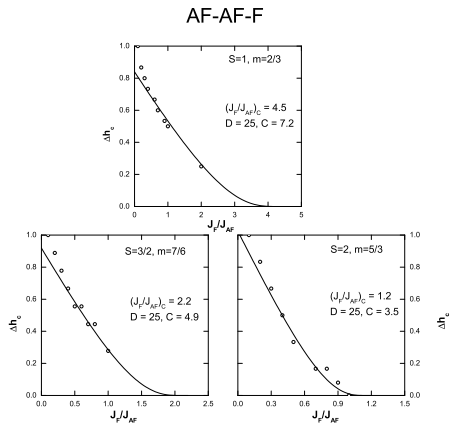}
\caption{ The plateau-non-plateau transition point $(J_{\rm F}/J_{\rm AF})_{\rm c}$ for
the AF-AF-F chain with $S=1$, $3/2$ and $2$, respectively, where $D,C$ are
constants in Eq. (\protect\ref{aaf-BKT}), and $\Delta h_{\rm c}(J_{\rm F}/J_{\rm AF})$
are scaled by the value $\Delta h_{\rm c}(0.1)$. }
\label{aaf-jc}
\end{figure}

As the width of the plateau might be extremely small near the transition
point, a direct estimation of the critical point from the numerical data may
be somewhat inaccurate. However, Eq. (\ref{aaf-BKT}) can be used to estimate
the plateau-non-plateau transition point $(J_{\rm F}/J_{\rm AF})_{\rm c}$ from the raw
numerical data, which yields $(J_{\rm F}/J_{\rm AF})_{\rm c}=4.5$ for $S=1$ and $m=2/3$, 
$2.2$ for $S=3/2$ and $m=7/6$, and $1.2$ for $S=2$ and $m=5/3$,
respectively, as shown in Fig. \ref{aaf-jc}. The larger the spin $S$ is, the
smaller the critical value of $(J_{\rm F}/J_{\rm AF})_{\rm c}$ is. It is also noticeable
that for a given $S$, the critical values for different plateaus are
slightly different. No fundamental difference between the systems with spin
integer and half-integer is observed.

\subsection{F-F-AF CHAIN}

Now we consider the case with $J=J_{\rm F}$, $J^{\prime }=J_{\rm AF}$, namely, the
system is a F-F-AF chain, as shown in Fig. \ref{chain}(b). It is an
antiferromagnet. In contrast to the AF-AF-F chain, when $J_{\rm AF}/J_{\rm F}$ is
much less than $1$, the ferromagnetic coupling is dominant; while $%
J_{\rm AF}/J_{\rm F}$ is greater than $1$, the antiferromagnetic interaction
predominates. We shall take the ratio $J_{\rm AF}/J_{\rm F}$ in the range of $%
[0.1,8] $ in this subsection.

\begin{figure}[tbp]
\vspace{0.5cm} \includegraphics[width = 0.75\linewidth,clip]{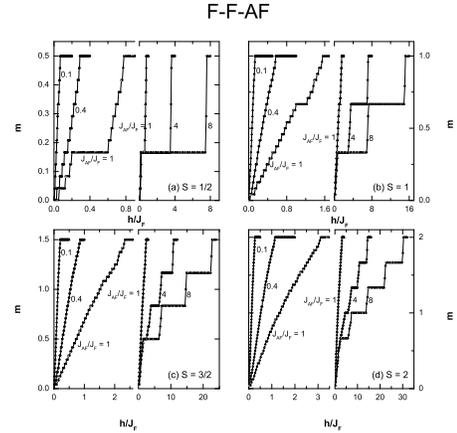}
\caption{Magnetization process of the F-F-AF chain with $S=1/2$, $1$, $3/2$
and $2$, respectively, where $h$ is the applied magnetic field, and $m$ is
the magnetization per site. $J = J_{\rm F} = 1$, $J^{\prime } = J_{\rm AF}$. }
\label{ffa-mh}
\end{figure}

Fig. \ref{ffa-mh} shows the magnetization process of the F-F-AF chain with
different spins. For $S=1/2$, we have observed the two plateaus at $m=1/6$
and $1/2$, consistent with the condition of Eq.(\ref{plateau}) with $n=3$.
For $S=1$, the three plateaus at $m=1/3$, $2/3$ and $1$ (saturation plateau)
are observed. For $S=3/2$, the four plateaus at $m=1/2$, $5/6$, $7/6$ and $%
3/2$ (saturation plateau) are seen, and for $S=2$, the five plateaus at $%
m=2/3$, $1$, $4/3$, $5/3$ and $2$ are obtained. It is seen that the smaller
the ratio $J_{\rm AF}/J_{\rm F}$ is, the less obvious the plateaus (except for the
saturation plateaus) are. In other words, when the AF coupling is
predominant in this system, the plateau is more obvious, and the width of
the plateau is much wider. In between the plateaus, the magnetization
increases rapidly with the magnetic field. From Fig. \ref{ffa-mh}, one may
find that, the plateaus at $m=0$ for $S=1$, $m=1/6$ for $S=3/2$, and $m=0$
and $1/3$ for $S=2$, which are also allowed by Eq.(\ref{plateau}) with $n=3$%
, are not observed. This is because Eq.(\ref{plateau}) is a necessary
condition for the occurrence of the magnetization plateaus. Therefore, the
number of the plateaus is still $(2S+1)$. As the present system is an
antiferromagnet, the magnetic curves look different to those of the AF-AF-F
chain. The saturation fields depend closely on the magnitude of the ratio $%
J_{\rm AF}/J_{\rm F}$, in contrast to the AF-AF-F chain. In addition, the similar
behaviors for the system with different spins are obtained, and no
fundamental difference of the properties of the system with spin integer and
half-integer is observed.

\begin{figure}[tbp]
\vspace{0.5cm} \includegraphics[width = 0.75\linewidth,clip]{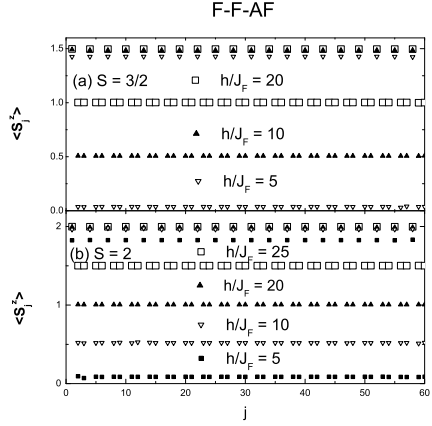}
\caption{ The spatial variation of $\langle S_{j}^{z}\rangle$ in the ground state for the
finite F-F-AF chain with length $L=60$ at $J_{\rm AF}/J_{\rm F}=8$. (a) For $S=3/2$,
the external field is $h/J_{\rm F}=5$, $10$, $20$, respectively. (b) For $S=2$,
the external field is $h/J_{\rm F}=5$, $10$, $20$ and $25$, respectively. }
\label{ffa-szj}
\end{figure}

Fig. \ref{ffa-szj} shows the behavior of $\langle S_{j}^{z}\rangle$ against $j$ for the
F-F-AF chain with spin $S=3/2$ and $2$, respectively. For both cases, the
ratio $J_{\rm AF}/J_{\rm F}=8$ is set. Fig. \ref{ffa-szj}(a) shows the case with $%
S=3/2$. For the external field $h/J_{\rm F}=5$, the spatial variation of 
$\langle S_{j}^{z}\rangle$ is $\{...,(0.0366,0.0365,1.4269),...\}$, giving rise to $m=1/2$%
. As the field is $h/J_{\rm F}=10$, the behavior of $\langle S_{j}^{z}\rangle$ against $j$
becomes $\{...,(0.5071,0.5071,1.4857),...\}$, resulting in $m=5/6$. For $%
h/J_{\rm F}=20$, the spin configuration is $\{...,(1.0018,1.0018,1.4964),...\}$,
giving $m=7/6$. Fig. \ref{ffa-szj}(b) shows the case with $S=2$. For $%
h/J_{\rm F}=5$, the expectation values $\langle S_{j}^{z}\rangle$ is $%
\{...,(0.0871,0.0863,1.8265),...\}$, leading to $m=2/3$. When $h/J_{\rm F}=10$,
the behavior of $\langle S_{j}^{z}\rangle$ versus $j$ becomes $%
\{...,(0.5181,0.5183,1.9636),...\}$, giving $m=1$; at $h/J_{\rm F}=20$, it
becomes $\{...,(1.0060,1.0060,1.9882),...\}$, yielding $m=4/3$; at $%
h/J_{\rm F}=25$, the spin configuration is $\{...,(1.5018,1.5018,1.9964),...\}$,
giving $m=5/3$. The $m$ values are just corresponding to the magnetization
plateaus. In addition, the spin configuration for the parameters above are stable
against the small local field $h_{\rm loc}/J_{\rm F}$=$0.1$ at $j=1$ site, which is
parallel to the external magnetic field $h$.

\begin{figure}[tbp]
\vspace{0.5cm} \includegraphics[width = 0.75\linewidth,clip]{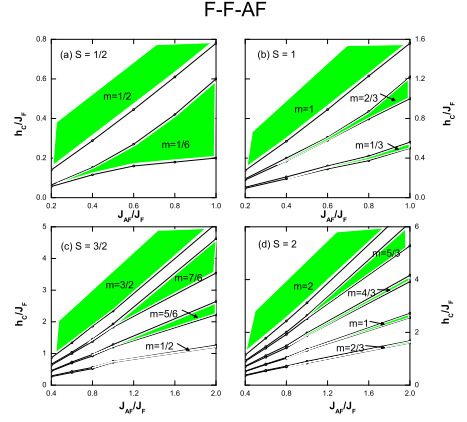}
\caption{ (Color online) Phase diagram in the $h_{\rm c}-J_{\rm AF}/J_{\rm F}$ plane for the
F-F-AF chain with $S=1/2$, $1$, $3/2$ and $2$, respectively. In the shaded
regions, the magnetization plateaus labeled with the quantized values of $m$
appear. }
\label{ffa-hc}
\end{figure}

Fig. \ref{ffa-hc} presents the phase diagram in the $h_{\rm c}-J_{\rm AF}/J_{\rm F}$
plane for the F-F-AF chain with spin $S=1/2$, $1$, $3/2$ and $2$,
respectively. In the shaded regions, the magnetization plateaus occur. At
the plateaus, the spin excitations from the ground state are gapful. When $%
J_{\rm AF}/J_{\rm F}$ increases beyond the critical value $(J_{\rm AF}/J_{\rm F})_{\rm c}$ at
which the plateau occurs, i.e. the AF coupling becomes more dominant, the
magnetization plateaus tend to appear. The plateau-non-plateau transition is
also expected to be of the BKT-type, and the width of the plateau $m=\Delta
h_{\rm c}$ near the transition point behaves as 
\begin{equation}
\mathrm{\Delta h_{\it c}=Dexp(-C/\sqrt{J_{\it AF}/J_{\it F}
         -(J_{\it AF}/J_{\it F})_{\it c}}) },
\label{ffa-BKT}
\end{equation}%
where $D,C$ are two constants \cite{FFACRI}.

\begin{figure}[tbp]
\vspace{0.5cm} \includegraphics[width = 0.75\linewidth,clip]{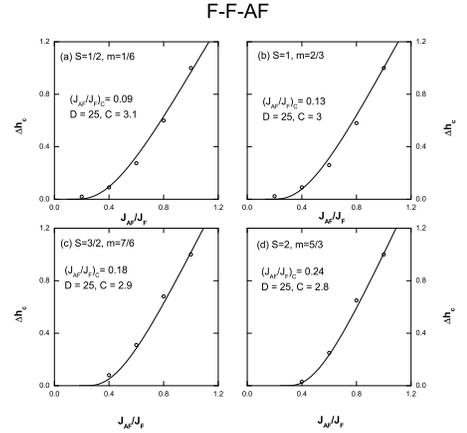}
\caption{ The plateau-non-plateau transition point $(J_{\rm AF}/J_{\rm F})_{\rm c}$ for
the F-F-AF chain with $S=1/2$, $1$, $3/2$ and $2$, respectively, where $D,C$
are constants in Eq. (\protect\ref{ffa-BKT}), and $\Delta
h_{\rm c}(J_{\rm AF}/J_{\rm F}) $ are scaled by the value $\Delta h_{\rm c}(1.0)$.}
\label{ffa-jc}
\end{figure}

As the width of the plateau might be extremely small near the transition
point, Eq. (\ref{ffa-BKT}) is used to estimate the plateau-non-plateau
transition point $(J_{\rm AF}/J_{\rm F})_{\rm c}$ from the raw numerical data. It gives $%
(J_{\rm AF}/J_{\rm F})_{\rm c}=0.09$ for $S=1/2$ and $m=1/6$, $0.13$ for $S=1$ and $m=2/3
$, $0.18$ for $S=3/2$ and $m=7/6$, and $0.24$ for $S=2$ and $m=5/3$,
respectively, as shown in Fig. \ref{ffa-jc}. It should be pointed out that
the breakdown of the magnetization plateau of $S=1/2$ F-F-AF chain has been
studied, with the estimation of the critical value $(J_{\rm AF}/J_{\rm F})_{\rm c}=0.065$%
\cite{FFACRI}. It can be seen that the larger the spin $S$ is, the greater
the value of $(J_{\rm AF}/J_{\rm F})_{\rm c}$ is. For a given $S$, the value of $%
(J_{\rm AF}/J_{\rm F})_{\rm c}$ for different plateaus is slightly different. In
addition, no fundamental difference between the F-F-AF systems with spin
integer and half-integer is also observed.

\section{SUMMARY AND DISCUSSION}

In this paper, by using the DMRG method we have numerically studied the
magnetic properties in the ground states of the J-J-J' trimerized quantum
Heisenberg chains with spin $S=1/2$, $1$, $3/2$ and $2$, respectively. The
two different cases are considered: (a) $J=J_{\rm AF}$ and $J^{\prime }=J_{\rm F}$,
i.e., the AF-AF-F trimerized ferrimagnetic chain. The magnetization plateaus
are observed when the AF coupling is dominated. The positions of these
plateaus are allowed by Eq.(\ref{plateau}) with $n=3$, and the number of the
plateaus is to be $(2S+1)$. For a certain spin $S$ and a plateau ($m\neq S$%
), the width of magnetization plateaus decreases with increasing the F
coupling, and becomes zero when the ratio $J_{\rm F}/J_{\rm AF} $ passes a critical
value, e.g. $(J_{\rm F}/J_{\rm AF})_{\rm c}=4.5$ for $S=1$ and $m=2/3 $, $2.2$ for $S=3/2
$ and $m=7/6$, and $1.2$ for $S=2$ and $m=5/3$, respectively. The larger the
spin $S$ is, the smaller the critical value of $(J_{\rm F}/J_{\rm AF})_{\rm c}$ is. For
a given $S$, the value of $(J_{\rm F}/J_{\rm AF})_{\rm c}$ for different plateaus is
slightly different. Furthermore, the saturation field dose not change with
the F coupling. (b) $J=J_{\rm F}$ and $J^{\prime }=J_{\rm AF}$, i.e., the F-F-AF
trimerized antiferromagnetic chain. The critical value in the
plateau-non-plateau transition is about $(J_{\rm AF}/J_{\rm F})_{\rm c}=0.09$ for $S=1/2$
and $m=1/6$, $0.13$ for $S=1$ and $m=2/3 $, $0.18$ for $S=3/2$ and $m=7/6$,
and $0.24$ for $S=2$ and $m=5/3$, respectively. For a given $S$, the value
of $(J_{\rm AF}/J_{\rm F})_{\rm c}$ for different plateaus is slightly different. We
would like to point out that though the ground state properties of the
F-F-AF chain are quite different from those of the AF-AF-F chain, as the
former is a antiferromagnet, while the latter is an ferrimagnet, the
magnetization plateaus in both cases tend to disappear when the
ferromagnetic coupling becomes more dominant.

As the width of the plateau might be extremely small near the transition
point, a direct estimation of the critical point from the numerical data may
be somewhat inaccurate, as mentioned above. However, near the transition
point the width of the plateau is expected to behave as $D{\rm exp}(-C/\sqrt{%
|J^{\prime }/J-(J^{\prime }/J)_{\rm c}|})$, where $D,C$ are two constants.
According to this property, $(J^{\prime }/J)_{\rm c}$ can be estimated from the
raw numerical data. Such an estimation is consistent with the previous
study, showing that our present calculations are reliable.

The appearance of the magnetization plateaus can be understood from the
spatial dependence of the averaged local magnetic moment.
For the AF-AF-F chain with spin $S=3/2$, as the ratio $J_{\rm F}/J_{\rm AF}=0.1$
and the external field $h/J_{\rm AF}=2$ are taken, the spatial variation
of $\langle S_{j}^{z}\rangle$ versus site $j$ in the ground state
follow the sequence such that $\{...,(1.2268,1.2268,-0.9535),...\}$,
resulting in the magnetization per site $m=
\sum_{j=1}^{N}(\langle S_{3j-2}^{z}\rangle+\langle S_{3j-1}^{z}\rangle
+\langle S_{3j}^{z}\rangle)/3N=1/2$,
corresponding to the plateau $m=1/2$. Similar cases are found when
the external field $h/J_{\rm AF}=3$ and $4$ are set. For the AF-AF-F chain
with spin $S=2$, F-F-AF chain with spin $S=3/2$ and $2$, the
similar spatial dependence of the averaged local magnetic moment are
observed. It should be noted that the spin configuration for the
parameters above are stable against the small local field
$h_{\rm loc}/J_{\rm AF}$=$0.1$ for AF-AF-F chain and $h_{\rm loc}/J_{\rm F}$=$0.1$ 
for F-F-AF chain, 
which is set at $j=1$ site and parallel to the
external magnetic field $h$.

By the spin-wave method starting from the fully polarized state, the
saturation magnetic field of the dimerized and quadrumerized AF chains can
be obtained analytically \cite{p-chain3}. Inspired by this method, we have
deduced an expression for the trimerized chains. The energy difference $%
\Delta E$ between the lowest energy with $M^{z}=(3N-1)S$ and that with $%
M^{z}=3NS$ is given by the lowest eigenvalue of the following matrix, 
\[
\left( 
\begin{array}{ccc}
-(J^{\prime }+1) & J^{\prime }e^{ik} & e^{-ik} \\ 
J^{\prime }e^{-ik} & -(J^{\prime }+1) & e^{ik} \\ 
e^{ik} & e^{-ik} & -2%
\end{array}%
\right) 
\]%
where $k$ is the wave number, $J=1$, $J^{\prime }$ and $S$ are the same
notation as mentioned above. We have found that the energy difference $%
\Delta E$ is the lowest when $k=\pi $. This gives rise to the saturation
field $H_{s}=-\Delta E=(3+2J^{\prime }+\sqrt{9-4J^{\prime }+4(J^{\prime
})^{2}})S/2$. When $J^{\prime }=1$, it becomes a simple AF chain, $H_{s}=4S$%
. When $J^{\prime }<0$ , it becomes an AF-AF-F chain, and $H_{s}$ is nearly
independent of $J^{\prime }$. These results are also in agreement with the
result of the quadrumerized chain\cite{p-chain3}. Most importantly, all
these analytic results are in agreement with our numerical results.

In trimerized Heisenberg chains, the total number of the plateaus is $2S+1$,
which implies that the OYA condition is only a necessary condition in
trimerized chains. This observation can be understood straightforward in the
case of ferrimagnet, i.e. the AF-AF-F chain. As shown in Fig. \ref{chain}%
(a), the possible minimum magnetization value is $m=S/3$, so the plateaus
whose magnetization value is smaller than $S/3$ cannot naturally appear. For
example, for $S=3/2$, the plateaus permitted by the OYA condition should be $%
m=1/6$, $1/2$, $5/6$, $6/7$ and $3/2$, respectively. As the plateau $m=1/6$
is smaller than the minimum magnetization of $S/3=1/2$, it does not appear.
Then, the number of the emerging plateaus is $4=2S+1$. However, this
argumentation seems not to apply to the case of antiferromagnet, i.e. the
F-F-AF chain. In Fig. \ref{chain}(b), the possible minimum magnetization
value is $m=0$. The plateaus $m=1/6$ of $S=3/2$ would be expected to emerge
during the magnetization, but they do not appear. The reason why the plateau 
$m=1/6$ of $S=3/2$ vanishes is under exploration. It is worthy of stressing
that for a certain spin $S$ the number of disappearance of plateaus is
smaller than that of the emergence of plateaus.

The obtained results of the two cases reveal that the degree of the
inhomogeneity of the couplings measured by the ratio $J^{\prime }/J$, not
just the period of the quantum spin chain indicated in Eq.(\ref{plateau}),
determines its ground-state properties. In addition, no fundamental
difference of the properties of the system with spin integer and
half-integer is observed. It should be noted that the magnetization plateaus
are only related to the AF coupling. In cases (a) and (b), when the F
coupling is much stronger than the AF one, there is no magnetization
plateaus at all, though the ratio of the couplings can still be large.
Clearly, these magnetization plateaus come from the quantum origin, that are
closely related to the competition between the exchange couplings. The
thermodynamic properties of this system are studied under the way.

\acknowledgments

This work is supported in part by the National Science Foundation of China
(Grant Nos. 90403036, 20490210, 10247002).

\end{document}